\documentclass[10pt,twocolumn]{article}

\usepackage{amsmath,amssymb,amsfonts}
\usepackage{algorithmic}
\usepackage{graphicx}
\usepackage{float}
\usepackage{multirow}
\usepackage{url}
\usepackage{adjustbox}
\usepackage{xcolor}
\usepackage{textcomp}
\usepackage[utf8]{inputenc}
\usepackage{longtable}
\usepackage[font=normalsize,labelfont=bf]{caption} 

\usepackage{geometry}
\usepackage{fancyhdr}
\usepackage{titlesec}
\usepackage{authblk} 

\titleformat{\section}{\large\bfseries}{\thesection}{1em}{}
\titleformat{\subsection}{\normalsize\bfseries}{\thesubsection}{1em}{}

\title{\textbf{I2I-Galip: Unsupervised Medical Image Translation Using Generative Adversarial CLIP}}
\author[]{\textbf{Yilmaz Korkmaz, Vishal M. Patel}}
\affil[]{Department of Electrical and Computer Engineering, Johns Hopkins University}
\affil[]{\{ykorkma1,vpatel36\}@jhu.edu}

\date{} 

\usepackage[pagebackref,breaklinks,colorlinks]{hyperref}

\usepackage[capitalize]{cleveref}
\crefname{section}{Sec.}{Secs.}
\Crefname{section}{Section}{Sections}
\Crefname{table}{Table}{Tables}
\crefname{table}{Tab.}{Tabs.}

\begin{document}
\def\SB#1{\textsubscript{#1}}

\maketitle              
\begin{abstract}
Unpaired image-to-image translation is a challenging task due to the absence of paired examples, which complicates learning the complex mappings between the distinct distributions of the source and target domains. One of the most commonly used approach for this task is CycleGAN which requires the training of a new pair of generator-discriminator networks for each domain pair. In this paper, we propose a new image-to-image translation framework named Image-to-Image-Generative-Adversarial-CLIP (I2I-Galip) where we utilize a pre-trained multi-model foundation model (i.e., CLIP) to mitigate the need of separate generator-discriminator pairs for each source-target mapping while achieving better and more efficient multi-domain translation. By utilizing the massive knowledge gathered during pre-training a foundation model, our approach makes use of a single lightweight generator network with $\approx$13M parameters for the multi-domain image translation task. Comprehensive experiments on translation performance in public MRI and CT datasets show the superior performance of the proposed framework over the existing approaches. Code will be available \href{https://github.com/yilmazkorkmaz1/I2I-Galip}{here}.
\end{abstract}

\section{Introduction}

Medical image translation is a challenging task because of significantly different domain distributions, necessitating the learning of very complex mappings between different imaging modalities \cite{roy2013atlas}.  Many supervised deep learning-based image translation methods have been proposed to address this problem \cite{dar2019image,jiang2023cola,armanious2020medgan}. However, these methods are limited due to the requirement of paired training data which might be challenging to acquire in real case scenarios. To overcome this constraint, various unsupervised image translation methods have been introduced for both general computer vision and medical imaging tasks \cite{dai2020multimodal,liu2017unsupervised,huang2018multimodal,ozbey2023unsupervised,han2021dual,yi2017dualgan,torbunov2023uvcgan}.  CycleGAN \cite{zhu2017unpaired} was one of the first approaches that proposed unpaired image translation which loosened the requirement for paired datasets by enforcing cycle-consistency among inverse translations. However, in the case of multiple modalities, CycleGAN introduces significant computational requirements as separate generator-discriminator pairs are required for each new modality. To mitigate the need of separate network pairs several multi-domain translation frameworks have been proposed \cite{choi2018stargan,choi2020stargan,huang2018multimodal,lee2018diverse}. Nonetheless, these methods generally lag in performance compared to uni-modal approaches.

More recently, a couple of text-driven diffusion based image-to-image translation frameworks have been proposed that integrate large vision-language pre-trained models as guidance \cite{tumanyan2023plug,rombach2022high,hertz2022prompt,kwon2022diffusion}, enabling robust translation across multiple domains. While these models provide zero-shot editing capabilities for various text conditions, they are limited in delivering fidelity necessary for the medical tasks. Moreover, these methods impose a significant computational burden due to the requirement for large denoiser backbones and extended inference times in their backward diffusion processes.


In this paper, we propose a cycle-consistent generative adversarial model to address the aforementioned limitations. Our model integrates BiomedCLIP (see \cref{biomedclip}), a pre-trained multi-modal vision-language model specifically trained in the medical domain, within a cycle-consistent feed-forward framework. By leveraging contrastive information from this large pre-trained network, we eliminate the need to train a new generator network for each translation task and reduce the requirement for large discriminators in feature extraction. Furthermore, our model enhances overall translation performance compared to existing unsupervised approaches in both single and multi-domain translation tasks. 

Our main contributions can be summarized as follows:
\begin{itemize}
    \item We introduce an adversarial framework for language-driven image translation for medical images.
    \item Our framework employs a novel CLIP driven cycle-consistent image translation model.
    \item Our model outperforms existing unsupervised baselines with a relatively lightweight backbone. Extensive experiments demonstrate its superior performance across various publicly available datasets from different modalities.
\end{itemize}

\section{Related Works}
\paragraph{Cycle-consistent image translation.}
Zhu et al. revolutionized the field of unsupervised image translation with their proposal of CycleGAN \cite{zhu2017unpaired}. Yi et al. proposed DualGAN \cite{yi2017dualgan} which is a concurrent work with CycleGAN offering the same cycle-consistency loss. Various studies followed the cycle-consistency constraint for more faithful translation in the unsupervised setting. Liu et al. proposed UNIT \cite{liu2017unsupervised} for uni-modal translation where a shared latent space is assumed between source and target modalities. Huang et al. proposed MUNIT \cite{huang2018multimodal} where UNIT's assumption of shared latent space is divided into content and style for multi-domain translation. Lee et al. \cite{lee2018diverse} introduced DRIT, which shares a similar approach to MUNIT by using disentangled content and attribute latents for multi-domain translation. Choi et al. proposed StarGANv1 \cite{choi2018stargan} and StarGANv2 \cite{choi2020stargan} where they utilized a separate style encoder network to generate distinct style codes to be used in generator for multi-domain translation. Perera et al. \cite{perera2018in2i} proposed an alternative method where they utilize multi-domain input modalities with a latent-consistency loss. Kim et al. proposed U-GAT-IT \cite{kim2019u} with an advanced generator equipped with adaptive layer instance normalization layers and attention. Torbunov et al. proposed UVCGan \cite{torbunov2023uvcgan} employing a pre-trained vision transformer as generator in a cycle-consistent framework for improved translation performance.

\paragraph{Text-guided image translation.}
Following the advancements in vision-language models \cite{radford2021learning} several text-guided unsupervised image translation methods proposed with or without cycle-consistency constraint. Park et al. proposed LANIT \cite{park2023lanit} where they use CLIP to generate pseudo labels for unlabeled images with a similar approach in Starganv2. Gal et al. proposed StyleGAN-NADA \cite{gal2022stylegan} for CLIP driven adaptation of Stylegan2 generator \cite{karras2020analyzing}. Patashnik et al. proposed StyleCLIP \cite{patashnik2021styleclip} where they invert source image to find its latent code for CLIP guided feature manipulation.

\paragraph{Diffusion-based image translation.}
More recently, building on the success of diffusion models in image generation, various unsupervised image translation methods utilizing diffusion-based backbones have been proposed. Zhao et al. proposed EGSDE \cite{zhao2022egsde} where they utilize energy-guided translation between diversely trained diffusion models. Özbey et al. proposed SynDiff \cite{ozbey2023unsupervised}, where they use multiple cycle-consistent diffusive and non-diffusive generators for improved translation performance. Kwon et al. proposed DiffuseIT \cite{kwon2022diffusion} and used pre-trained vision transformers as guidance in image manipulation. Tumanyan et al. \cite{tumanyan2023plug} offered a plug and play framework to adapt pre-trained text-to-image diffusion models in image translation. Zhan et al. proposed MedM2G \cite{zhan2024medm2g}, where they proposed a unified multi-modal diffusive framework for text to image, image to text synthesis and image translation tasks. Parmar et al. \cite{parmar2024one} proposed a one step diffusion model for unsupervised image translation adapting the pre-trained latent diffusion weights. Liu et al. \cite{liu2024ladiffgan} offered an adversarial network utilizing diffusion supervision in latent space. 

Our approach shares similarities with MedM2G \cite{zhan2024medm2g} in employing a multi-modal text-guided framework for image translation. However, our model is over an order of magnitude smaller, leveraging a feed-forward generative adversarial network architecture and enforcing cycle-consistency across translations. We also incorporate common loss terms with DiffuseIT \cite{kwon2022diffusion}, utilizing CLS tokens from pre-trained vision transformers for semantically meaningful information extraction. Nonetheless, our approach differs in its use of cycle-consistency and the feed forward generative adversarial methodology adopted. We named our method in reference to the text-to-image generative adversarial model Galip \cite{tao2023galip}. However, apart from the CLIP based feature extraction utilized for the Discriminator, our method does not share further similarities with Galip in terms of architecture or training methodology.

\section{Background}
\subsection{Cycle-Consistent Generative Adversarial Networks (CycleGAN)}

CycleGAN \cite{zhu2017unpaired} models the unpaired image translation problem between domain A and B using two translators. First, two translators   (G : $A$ → $B$) and  (F : $B$ → $A$) are defined. Then  G and F are forced to be inverses of each other, thus making both mappings to be approximately bijections.  CycleGAN achieves remarkable performance using this cycle-consistency combined with the adversarial loss which encourages $F (G(X_A)) \approx X_A$ and $G(F (X_B)) \approx X_B$.

\subsection{BiomedCLIP}
\label{biomedclip}
In this paper, we utilize BiomedCLIP \cite{zhang2023large} as our pre-trained vision-language model. BiomedCLIP is trained on PMC-15M dataset using pairs of figures and captions from biomedical research articles in PubMed Central and outperforms other medical vision-language models in various tasks \cite{zhang2023large}. BiomedCLIP utilizes a ViT-B \cite{dosovitskiy2020image} based image encoder while utilizing PubMedBERT \cite{gu2021domain} for the text embeddings.

\section{Methodology}
\subsection{I2I-Galip}

\begin{figure*}[!t]
\includegraphics[width=1.0\textwidth]{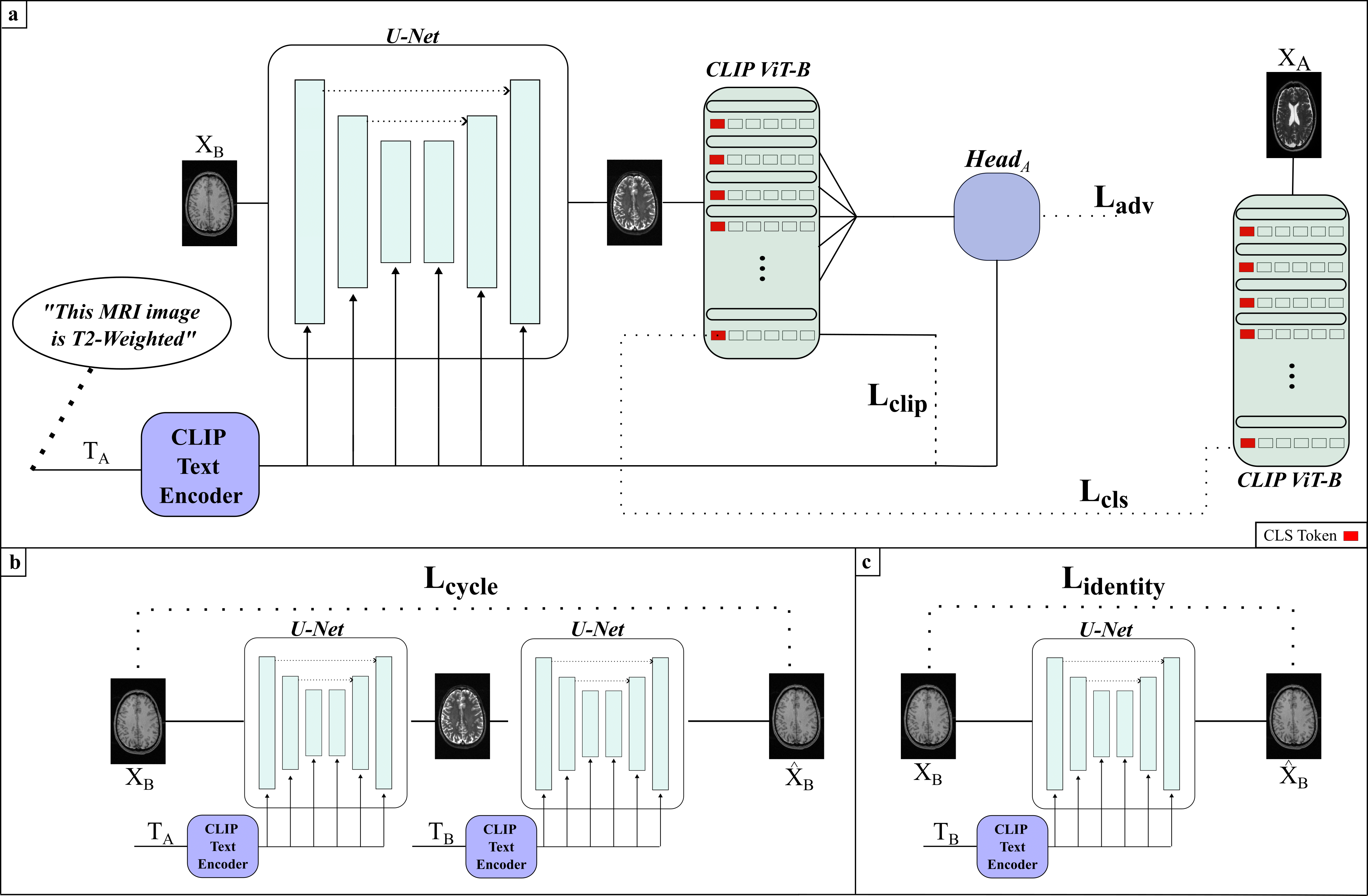}
\caption{Training scheme and overall model architecture of I2I-Galip is illustrated when input image is from domain B. Part a illustrates $L_{clip}$, ${L}_{cls}$ and ${L}_{adv}$ losses along with U-Net based generator, discriminator head and BiomedCLIP's ViT-B. Part b and c shows the definition of ${L}_{cycle}$ and ${L}_{identity}$ losses respectively. BiomedCLIP's ViT-B and text encoder parameters are frozen during training. "This MRI Image is T\SB{2}-weighted" corresponds to a sample caption used in T\SB{1} to T\SB{2} translation.}
\label{fig:fig_1}
\end{figure*}

We design a lightweight generator network which is a very thin variant of the latent diffusion U-Net \cite{rombach2022high} (with only $\approx13$M parameters). Our discriminator network is using the projections of intermediate Vision Transformer (ViT) features as input, adapted from text-to-image model Stylegan-T \cite{sauer2023stylegan}. This discriminator design  allow  us to utilize the output of different layers in BiomedCLIP's ViT, capturing different level of details. We modified this design by dividing the discriminator heads into distinct sets, tailored specifically for a target translation domain. We also utilize BiomedCLIP's text encoder to generate target text embeddings using captions for each modality (see \cref{fig:fig_1}a), which  controls the generated image features via cross-attention transformers while serving as a regularizer in the training (shown in \cref{fig:fig_1}a). Overall training objective for the generator can be expressed as follows
\begin{equation}
\begin{aligned}
    \mathcal{L}_{total} = \, & \lambda_{cycle} \cdot \mathcal{L}_{cycle} + \lambda_{adv} \cdot \mathcal{L}_{adv_G} \\
    & + \lambda_{clip} \cdot \mathcal{L}_{clip} + \lambda_{cls} \cdot \mathcal{L}_{cls} \\
    & + \lambda_{identity} \cdot \mathcal{L}_{identity},
\end{aligned}
\end{equation}

where $\lambda_{cycle}, \lambda_{adv}, \lambda_{clip}, \lambda_{cls}, \lambda_{identity} $ are coefficients to control the contribution from each loss. We denote the loss associated with the discriminator as $\mathcal{L}_{adv_D}$, apart from that other loss terms are utilized in generator training. In what follows, we describe   each of these loss terms in detail.
\begin{enumerate}
\item \textbf{Adversarial Loss}: By leveraging intermediate features from the ViT, direct feature extraction from images becomes unnecessary, enabling the use of lightweight discriminator heads for each feature level. We utilize the least squares GAN loss \cite{mao2017least} to enhance the stability of training instead of Hinge loss used in StyleGAN-T, which can be defined as follows
\begin{equation}
    \begin{aligned}
        \mathcal{L}_{adv_G} = & \mathbb{E}[\left(Head_{A}(E_{X_A}^{out}, E_{T_A}) - 1\right)^2] \\
        & + \mathbb{E}[\left(Head_{B}(E_{X_B}^{out},E_{T_B}) - 1\right)^2],
    \end{aligned}
\end{equation}
    
\begin{equation}
    \begin{aligned}
        \mathcal{L}_{adv_D} = & \mathbb{E}\left[\left(Head_{A}(E_{X_A}^{input}, E_{T_A}) - 1\right)^2 \right. \\
        & \left. + \left(Head_{A}(E_{X_A}^{out}, E_{T_A})\right)^2\right] \\
        & + \mathbb{E}\left[\left(Head_{B}(E_{X_B}^{input}, E_{T_B}) - 1\right)^2 \right. \\
        & \left. + \left(Head_{B}(E_{X_B}^{out}, E_{T_B})\right)^2\right]
    \end{aligned}
\end{equation}

    where $Head_{A}$ and $Head_{B}$ corresponds to the discriminator heads allocated for the specific target domain, $E_{T_A}$ and $E_{T_B}$ are corresponding text encodings (i.e., text encodings of captions for each domain), $E_{X_A}^{input}$, $E_{X_B}^{input}$, $E_{X_A}^{out}$ and $E_{X_B}^{out}$ are feature maps from ViT for input and generated images from domain A and B, respectively. We describe how these terms are generated as follows
    \begin{align}
        E_{X_A}^{input}=ViT_{int}(X_A^{input}) \\
        E_{X_B}^{input}=ViT_{int}(X_B^{input}) \\
        E_{X_A}^{out}=ViT_{int}(G(X_B^{input}, {T_A}, Y_A)) \\
        E_{X_B}^{out}=ViT_{int}(G(X_A^{input}, {T_B}, Y_B)) \\
    \end{align}
    where $ViT_{int}$ corresponds to intermediate layers of ViT, $X_A^{input}$ and $X_B^{input}$ are unpaired input images, $Y_A$ and $Y_B$ are binary class labels for domain A and B, respectively.

    \item \textbf{Cycle Loss}: We enforce cycle-consistency, shown in \cref{fig:fig_1}b, to encourage more faithful translation between source and target domains for each pair which is defined as follows
    \begin{equation}
    \begin{aligned}
    \mathcal{L}_{\text{cycle}} = & \mathbb{E}\left[\|X_B^{input} - G(X_A^{input}, E_{T_B}, Y_B)\|_1\right] \\ 
     & + \mathbb{E}\left[\|X_A^{input} - G(X_B^{input}, E_{T_A}, Y_A)\|_1\right].
    \end{aligned}
    \end{equation}
    \item \textbf{CLIP Guidance Loss}: We minimize cosine distance between the text encoding corresponds to the caption of target domain (e.g., "This MRI image is T\SB{1}-Weighted" or "This is pelvic CT") and the encoding from ViT for the generated images to enable the utilization of CLIP's joint embedding space similarly with \cite{patashnik2021styleclip}, which can be defined as follows
    \begin{equation}
    \mathcal{L}_{\text{clip}} = -\frac{\langle E_{T_A},E_{X_A}^{out^{last}}  \rangle}{\|E_{T_A}\| \cdot \|E_{X_A}^{out^{last}}\|} - \frac{\langle E_{T_B},E_{X_B}^{out^{last}}  \rangle}{\|E_{T_B}\| \cdot \|E_{X_B}^{out^{last}}\|},
\end{equation}
    where $E_{X_A}^{out^{last}}$, $E_{X_B}^{out^{last}}$ are the image encodings from the last layer of ViT for the generated images from domain A and B, respectively. Generally, $\mathcal{L}_{clip}$ is dominated by cycle and adversarial losses giving comparingly small benefits (see \cref{sec:ablation} for details).

    \item \textbf{CLIP Encoding Loss}: The CLS tokens in the final layers of vision transformers are recognized for containing semantically rich information, as highlighted by \cite{tumanyan2022splicing}, which is typically leveraged for downstream classification tasks and shown to be beneficial in image translation \cite{kwon2022diffusion}. Therefore,  we enforce cosine similarity between the CLS tokens in the ViT for the generated and target domain's images to enforce semantic similarity among these images, which can be written as follows
\begin{equation}
    \mathcal{L}_{\text{cls}} = -\frac{\langle cls_{X_A}^{input} , cls_{X_A}^{out}  \rangle}{\|cls_{X_A}^{input}\| \cdot \|cls_{X_A}^{out}\|} - \frac{\langle cls_{X_B}^{input} , cls_{X_B}^{out}  \rangle}{\|cls_{X_B}^{input}\| \cdot \|cls_{X_B}^{out}\|},
\end{equation}
    where $cls_{X_A}^{input}$ and $cls_{X_B}^{input}$ are the CLS tokens in the last layers of ViT for input images from domain A and B respectively. $cls_{X_A}^{out}$ and $cls_{X_B}^{out}$ are corresponding CLS tokens for the generated images.
    
    \item \textbf{Identity Loss}: Identity loss is found to be beneficial to maintain source image structure in translation by enforcing the pixel-level equality when target and source domains match \cite{zhu2017unpaired}. We enforce it via using same labels and text embeddings corresponding to the input image domain (see \cref{fig:fig_1}c). It can be defined using our framework as follows
\begin{equation}
\begin{aligned}
        \mathcal{L}_{\text{identity}} = &\mathbb{E}\left[\|X_A^{input} - G(X_A^{input}, E_{T_A}, Y_A))\|_1\right] \\
        & + \mathbb{E}\left[\|X_B^{input} - G(X_B^{input}, E_{T_B}, Y_B)\|_1\right].
\end{aligned}
\end{equation}
\end{enumerate}

\begin{figure*}[!t]
\includegraphics[width=1\textwidth]{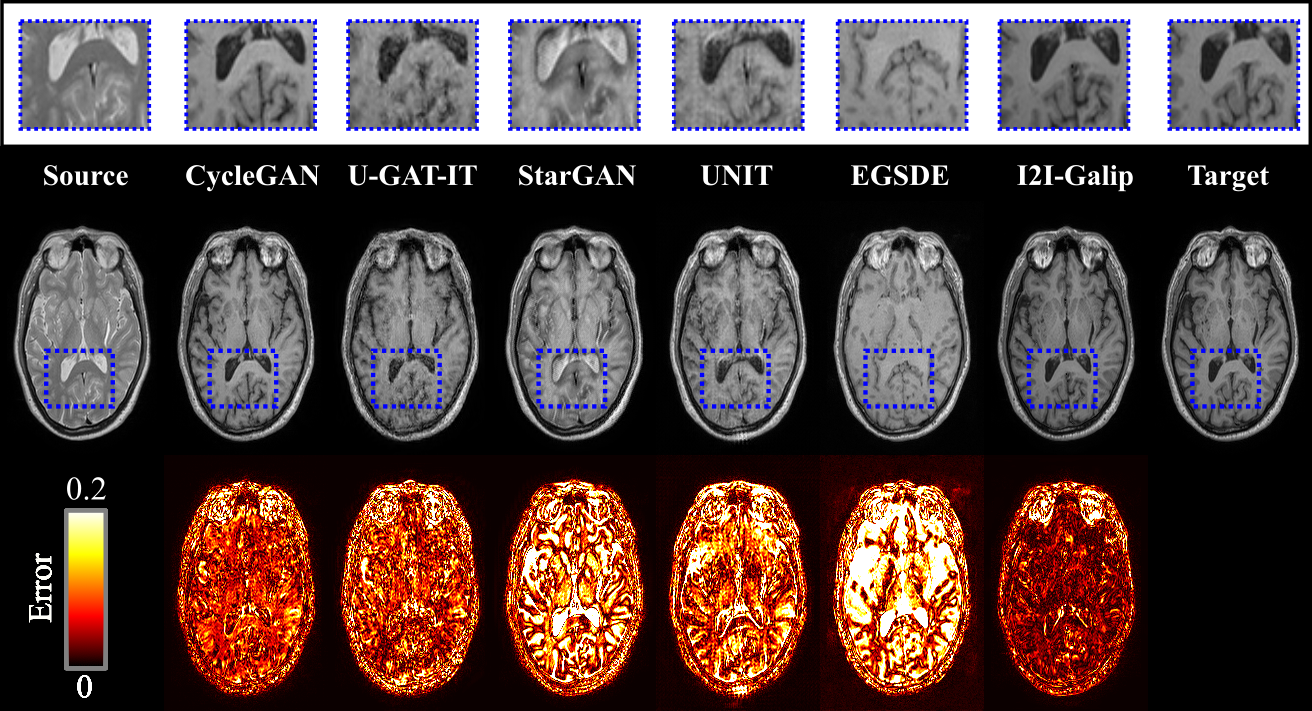}
\caption{Multi-domain translation illustrations from PD to T\SB{1}-weighted image in IXI dataset. Accompanying this are error maps and magnified sections, positioned below and above each translation, respectively.}
\label{fig:pd_to_t1}
\end{figure*}

\begin{figure*}[!t]
\includegraphics[width=1\textwidth]{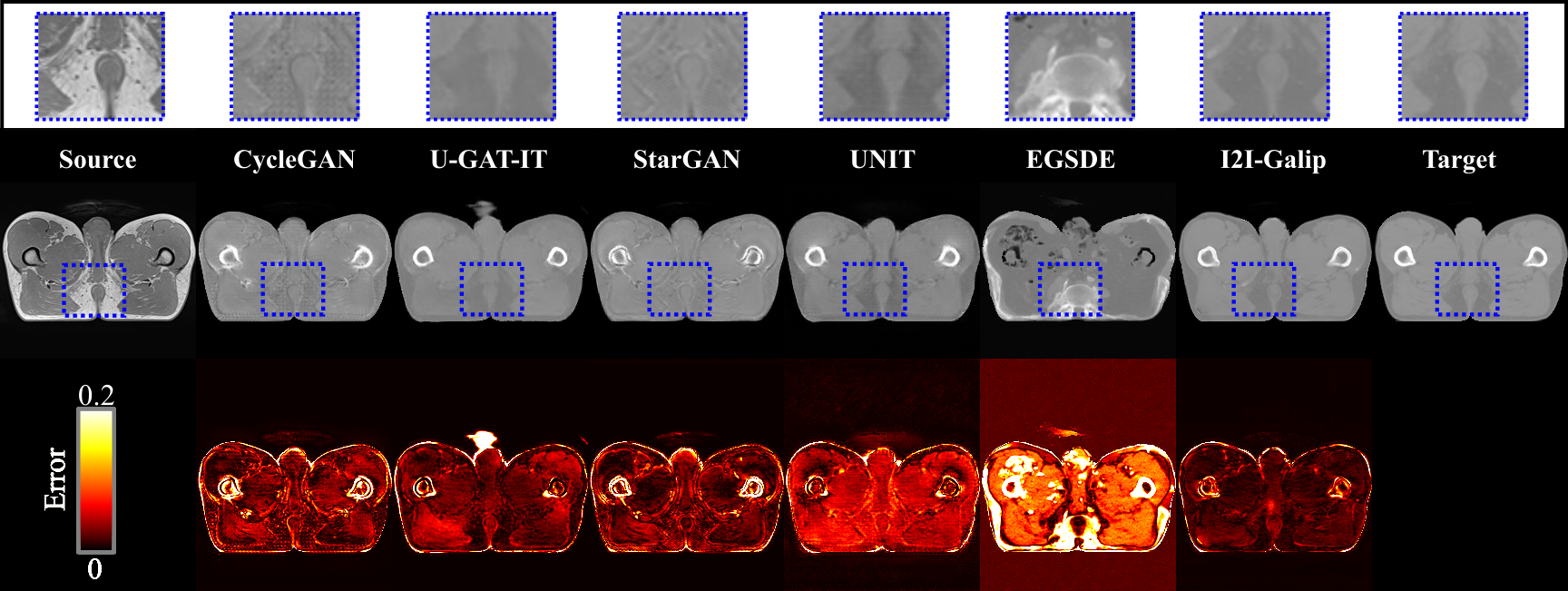}
\caption{Single-domain translation from T\SB{1}-weighted Pelvic MRI to CT images. Accompanying this are error maps and magnified sections, positioned below and above each translation, respectively.}
\label{fig:mri_to_ct}
\end{figure*}

\begin{table*}[]
\centering
\caption{Model complexities are illustrated using total number of parameters for each competing method. The third row indicates the number of required generator and discriminator networks, given the specified number of domain. $T$ and P(.) represents the number of domains for a multi-modal translation problem and permutation operator respectively. Total parameters are calculated for a representative case where $T=4$.}
\begin{tabular}{|c|c|c|c|c|c|c|}
\hline
Network/Model & I2I-Galip & CycleGAN       & U-GAT-IT       & StarGAN & UNIT           & EGSDE  \\ \hline
Generator (G)     & 13.2M        & 11.3M          & 278.9M         & 8.4M    & 5.4M + 5.4M    & 164M   \\ \hline
Discriminator (D) & 23.9M        & 2.7M           & 56.4M          & 44.8M   & 2.8M           & 0      \\ \hline
Times (G, D)         & 1, T      & P(T,2), P(T,2) & P(T,2), P(T,2) & 1, 1    & P(T,2), P(T,2) & T, 0   \\ \hline
Total         & 108.8M        & 169.5M          & 4023.6M        & 53.2M   & 162.2M          & 657.2M \\ \hline
\end{tabular}
\label{tab:model_comp}
\end{table*}

\begin{table*}[]
\centering
\caption{Multi-domain image translation results in IXI dataset. T\SB{1}-, T\SB{2}- and PD-weighted images are considered.}
\label{tab:my-table}
\resizebox{\textwidth}{!}{%
\begin{tabular}{|c|c|c|c|c|c|c|}
\hline
One-to-one task         & T\SB{1}-\textgreater{}T\SB{2}    & T\SB{2}-\textgreater{}T\SB{1}    & T\SB{2}-\textgreater{}PD    & PD-\textgreater{}T\SB{2}    & T\SB{1}-\textgreater{}PD    & PD-\textgreater{}T\SB{1}    \\ \hline
IXI                     & PSNR $\vert$ SSIM            & PSNR $\vert$ SSIM            & PSNR $\vert$ SSIM            & PSNR $\vert$ SSIM            & PSNR $\vert$ SSIM            & PSNR $\vert$ SSIM            \\ \hline
I2I-Galip-M   & 27.22 $\vert$ 90.18 & 27.30 $\vert$ 90.86 &\textbf{32.34} $\vert$ \textbf{95.74} &\textbf{33.12} $\vert$ 95.39 & 26.76 $\vert$ 90.75 & 27.70 $\vert$ \textbf{91.20} \\ \hline
I2I-Galip-S   & \textbf{27.47}$ \vert$ \textbf{90.54} & \textbf{27.33} $\vert$ \textbf{91.06} & 32.11 $\vert$ 95.65 & 32.87 $\vert$ \textbf{95.62} & \textbf{26.99} $\vert$ \textbf{90.80} & \textbf{27.75}$ \vert$ 91.07  \\ \hline
CycleGAN           & 26.10 $\vert$ 87.36          & 26.31 $\vert$ 88.51          & 27.43 $\vert$ 93.68          & 31.07 $\vert$ 93.81          & 24.56 $\vert$ 88.26          & 25.91 $\vert$ 89.47          \\ \hline
U-GAT-IT          & 24.44 $\vert$ 86.19          & 24.51 $\vert$ 86.85          & 26.81 $\vert$ 91.39          & 29.03 $\vert$ 92.11          & 22.98 $\vert$ 85.16          & 24.83 $\vert$ 87.44          \\ \hline
StarGAN  & 20.96 $\vert$ 71.40          & 21.00 $\vert$ 71.44          & 26.18 $\vert$ 91.80          & 27.33 $\vert$ 91.52          & 21.69 $\vert$ 72.92          & 21.51 $\vert$ 72.89          \\ \hline
UNIT &   23.59  $\vert$ 84.40                     &      24.76  $\vert$ 86.63                  &  25.22 $\vert$ 91.42                       &            29.10 $\vert$ 93.30            &    23.20 $\vert$ 86.00                    &           23.50 $\vert$ 80.05             \\ \hline
EGSDE        &   16.93 $\vert$ 53.32                     &      17.44 $\vert$ 57.54                  &      17.98 $\vert$ 75.93                  &     16.40 $\vert$ 57.55                    &           19.70 $\vert$ 71.21             &     19.71 $\vert$ 59.73                   \\ \hline
\end{tabular}%
}
\label{tab:ixi}
\end{table*}

\subsection{Datasets}
We conduct experiments on the following datasets to demonstrate the performance of our approach. 
\begin{enumerate}
    \item \textbf{IXI}: Translation performance demonstrated in a single-coil brain MRI dataset from (http://brain-development.org/ixi-dataset/). T\SB{1}-, T\SB{2}- and PD-weighted acquisitions are considered. In IXI, 25 subjects are used for training, 5 for validation and 10 for testing.
    \item \textbf{CT-MRI}: Translation performance demonstrated in pelvic T\SB{1}- and T\SB{2}-weighted MRI and CT data from \cite{nyholm2018mr}. In CT-MRI dataset, 9 subjects are used for training, 1 for validation and 4 for testing. 
\end{enumerate}

We consider the IXI dataset in both multi-domain and single-domain translation contexts. In the multi-domain scenario, we use a single network for all translation tasks, whereas in the single-domain scenario, we utilize distinct networks for each individual task. On the other hand, CT-MRI dataset only allows us to use single-domain translation context.

\begin{figure*}[!t]
\includegraphics[width=1\textwidth]{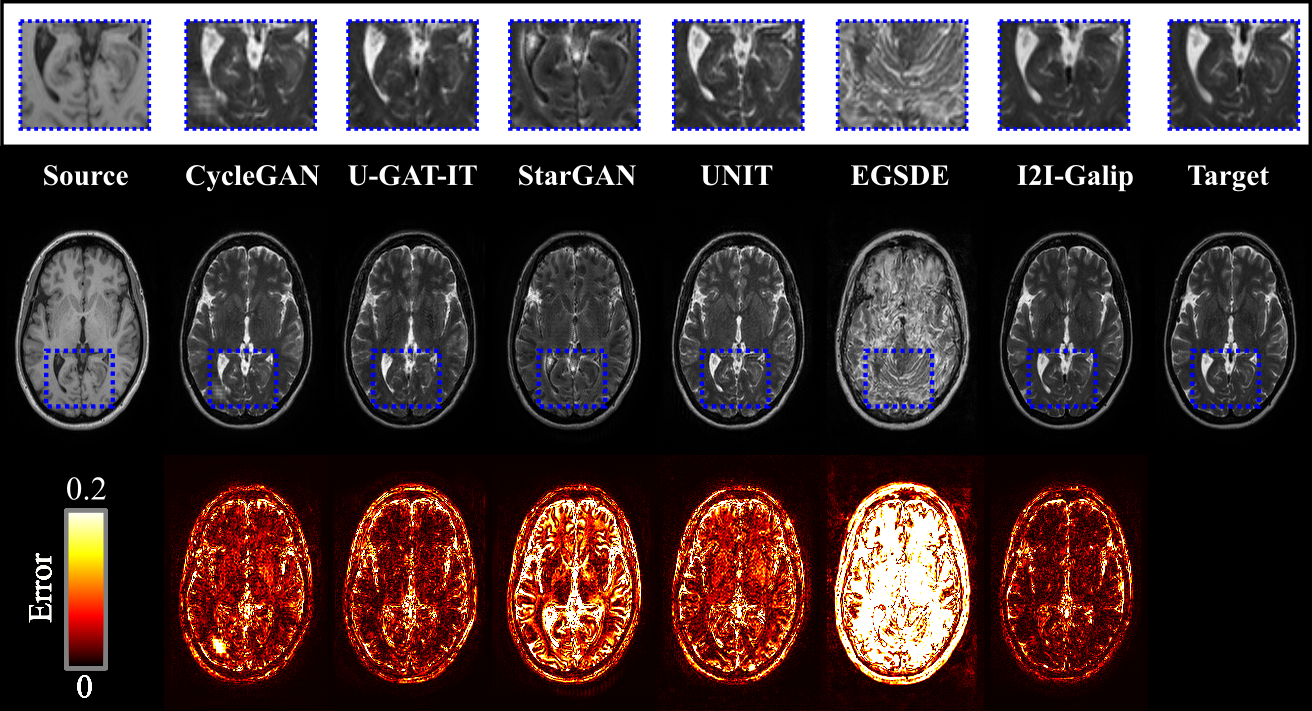}
\caption{Multi-domain translation illustrations from T\SB{1}-weighted to T\SB{2}-weighted image in IXI dataset. Accompanying this are error maps and magnified sections, positioned below and above each translation, respectively.}
\label{fig:fig_t1_to_t2}
\end{figure*}

\subsection{Competing Methods}
We illustrate the model complexities using the number of parameters in each competing method in the \cref{tab:model_comp}. A single NVIDIA RTX A5000 GPU with PyTorch framework is utilized in all experiments.
\begin{enumerate}
    \item \textbf{I2I-Galip}: Our model is trained with Adam optimizer with an initial learning rate set at 0.0002, which is linearly decreased to 0 after the $50th$ epoch. Number of discriminator head sets are determined according to the number of domains in the translation problem, where for IXI it is 3, and 2 for CT-MRI. We utilize hyperparameters 10, 1, 1, 1, 1 for $\lambda_{cycle}$, $\lambda_{adv}$, $\lambda_{cls}$, $\lambda_{clip}$, and $\lambda_{identity}$ respectively.
    \item \textbf{CycleGAN}: Cycle-consistent generative adversarial model is considered \cite{zhu2017unpaired}. 
    The Adam optimizer is utilized for training with an initial learning rate set at 0.0002, which linearly decreased to 0 after the $50th$ epoch. The training process spans a total of 100 epochs. Weights for adversarial, cycle, identity losses are selected as 1, 10, 0.5 respectively.
    \item \textbf{U-GAT-IT}: An attention guided GAN model with adaptive layer-instance normalization designed for unsupervised image translation is considered \cite{kim2019u}. Adam optimizer is utilized for training with a learning rate of 0.0001. Training lasts for 100 epochs. Weights for adversarial, cycle, identity and CAM losses are selected as 1, 10, 10, and 1000 respectively.
    \item \textbf{StarGAN}: A unified unsupervised image translation GAN model is considered \cite{choi2018stargan}. Adam optimizer is used for training with a learning rate of 0.0001. Training length is 100 epochs. Weights for domain classification loss, reconstruction loss and gradient penalty are selected as 1, 10, 10 respectively. A single StarGAN model is trained and tested for all domain pairs in each dataset.
    \item \textbf{UNIT}: An unsupervised GAN model designed for unsupervised image translation is considered \cite{liu2017unsupervised}. Adam optimizer is utilized for training with a learning rate of 0.0001 for 100 epochs. Weights for adversarial, image, style, and content reconstruction losses are selected as 1, 10, 1, 1 respectively.
    \item \textbf{EGSDE}: A diffusion based unpaired image translation model is considered \cite{zhao2022egsde}. Seperate DDPM models are trained for each translation domain to be utilized in EGSDE model. 500,000 diffusion steps are used for training of the DDPMs and T is selected as 150 to maintain source structure, and cross-validated weight parameters $\lambda_s$ and $\lambda_i$ are selected as $1 \times 10^{-7}$ and 10.
\end{enumerate}

\begin{figure*}[!t]
\includegraphics[width=1\textwidth]{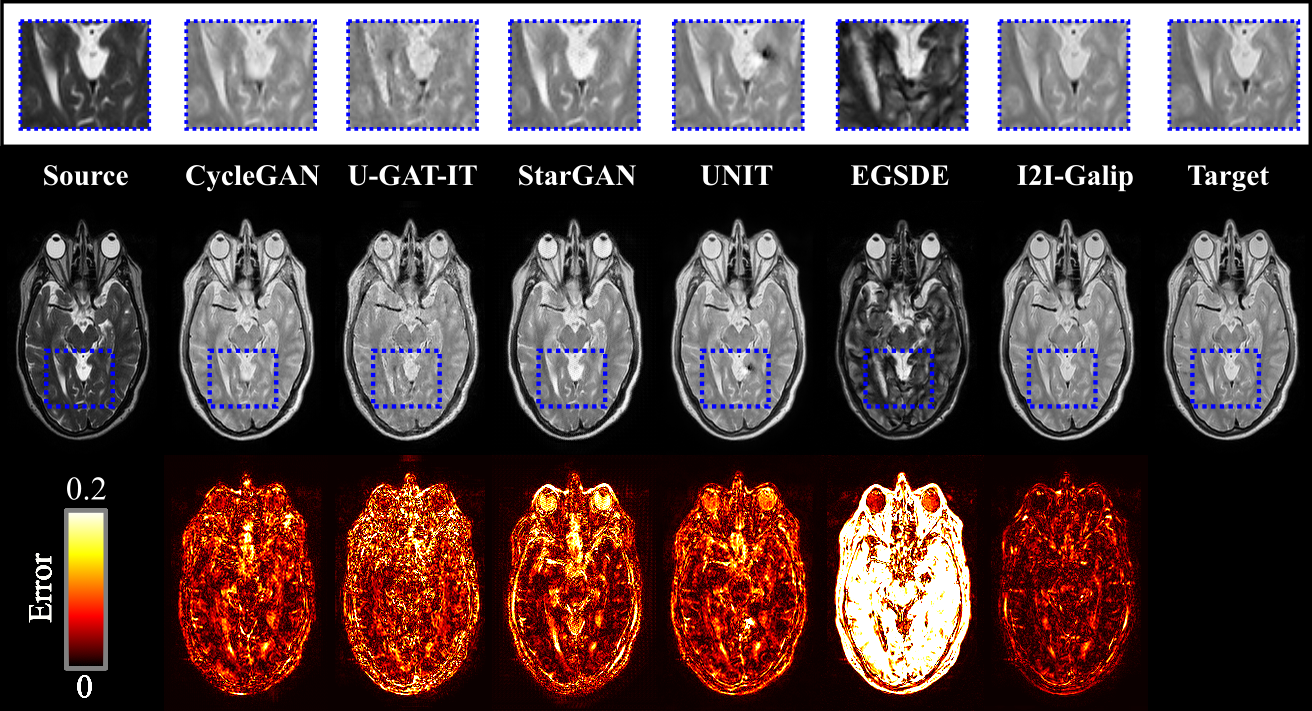}
\caption{Multi-domain translation illustrations from T\SB{2}-weighted to PD image in IXI dataset. Accompanying this are error maps and magnified sections, positioned below and above each translation, respectively.}
\label{fig:fig_t2_to_pd}
\end{figure*}

\section{Results}
\begin{table}[]
\centering
\caption{Single-domain image translation results in CT-MRI dataset for T\SB{1}- and T\SB{2}-weighted images.}
\label{tab:ct_mri}
\resizebox{0.35\textwidth}{!}{%
\begin{tabular}{|c|cc|cc|}
\hline
             & \multicolumn{2}{c|}{T\SB{1}-\textgreater{}CT} & \multicolumn{2}{c|}{T\SB{2}-\textgreater{}CT} \\ \hline
             & \multicolumn{1}{c|}{PSNR}     & SSIM     & \multicolumn{1}{c|}{PSNR}     & SSIM     \\ \hline
I2I-Galip & \multicolumn{1}{c|}{\textbf{26.13}}    & \textbf{90.86}    & \multicolumn{1}{c|}{27.08}    & \textbf{91.30}    \\ \hline
CycleGAN     & \multicolumn{1}{c|}{24.55}    & 78.63    & \multicolumn{1}{c|}{\textbf{27.39}}    & 89.92    \\ \hline
U-GAT-IT     & \multicolumn{1}{c|}{25.79}    & 89.34    & \multicolumn{1}{c|}{26.01}    & 87.48    \\ \hline
StarGAN      & \multicolumn{1}{c|}{25.00}    & 79.96    & \multicolumn{1}{c|}{24.67}    & 78.37    \\ \hline
UNIT         & \multicolumn{1}{c|}{26.02}    & 79.22    & \multicolumn{1}{c|}{25.15}    & 75.30    \\ \hline
EGSDE        & \multicolumn{1}{c|}{19.03}    & 74.63    & \multicolumn{1}{c|}{14.74}    & 66.67    \\ \hline
\end{tabular}}
\end{table}

We used Peak-Signal-to-Noise-Ratio (PSNR, dB) and Structural Similarity Index Measure (SSIM, \%) to compare the translation performances of competing methods. Results are presented for both single- and multi-domain case in IXI for I2I-Galip to show the effectiveness of the proposed approach for both cases. CT-MRI results are presented as the single-domain translation. I2I-Galip-S (Single), CycleGAN, U-GAT-IT, UNIT are separately trained for all possible domain pairs while I2I-Galip-M (Multi) and  StarGAN are trained once per dataset. EGSDE is a training free method for image translation but it requires separately trained diffusion models for each target domain. \cref{tab:ixi} and \cref{tab:ct_mri} shows the translation performance in IXI and CT-MRI datasets respectively. We show the corresponding translated images for each competing methods for distinct translation tasks in \cref{fig:pd_to_t1}, \cref{fig:mri_to_ct}, \cref{fig:fig_t1_to_t2} and \cref{fig:fig_t2_to_pd}. Best performances are highlighted as bold in each table for each metric. Overall, I2I-Galip-M yields 2.17dB better PSNR and over 2\% better SSIM than the second best competing method on average in IXI. I2I-Galip-S yields 0.10 dB better PSNR and 1.52\% better SSIM in T\SB{1} to CT, while providing 1.38\% better SSIM in T\SB{2} to CT task. Our method excels in capturing high-frequency details more effectively than competing approaches, providing superior image clarity and precision. Unlike other methods, it does not suffer from the noise artifacts that often degrade the quality of the output, leading to more accurate and visually appealing results even with its low computational budget as shown in \cref{tab:model_comp}. 

\subsection{Ablation Studies}
\label{sec:ablation}
We illustrate the effect of individual loss components in the proposed model for single- and multi-domain cases in  \cref{tab:ablation}. We observe that the most of the performance gain comes from adversarial and cycle losses although clip guidance, identity and clip encoding losses are beneficial especially in the multi-domain setting. In a single-domain setting, the adversarial loss tends to dominate, thereby reducing the influence of other loss terms. We discuss the reasons behind these results in \cref{sec:limitations}.
\begin{table}[]
\centering
\caption{Single- and multi-domain ablation results in IXI dataset. PSNR and SSIM values are averaged across the whole test set.}
\resizebox{0.35\textwidth}{!}{%
\begin{tabular}{|c|cc|cc|}
\hline
             & \multicolumn{2}{c|}{I2I-Galip-S}                          & \multicolumn{2}{c|}{I2I-Galip-M}                           \\ \hline
             & \multicolumn{2}{c|}{IXI}                             & \multicolumn{2}{c|}{IXI}                             \\ \hline
             & \multicolumn{1}{c|}{PSNR}           & SSIM           & \multicolumn{1}{c|}{PSNR}           & SSIM           \\ \hline
Proposed     & \multicolumn{1}{c|}{\textbf{29.09}} & \textbf{92.48} & \multicolumn{1}{c|}{\textbf{29.07}} & \textbf{92.35} \\ \hline
$\lambda_{adv}=0$   & \multicolumn{1}{c|}{19.80}          &60.80          & \multicolumn{1}{c|}{18.62}          & 49.05          \\ \hline
$\lambda_{cls}=0$   & \multicolumn{1}{c|}{29.00}          & 92.26          & \multicolumn{1}{c|}{28.88}          & 91.99          \\ \hline
$\lambda_{cycle}=0$ & \multicolumn{1}{c|}{27.91}          & 90.93          & \multicolumn{1}{c|}{28.08}          & 90.88          \\ \hline
$\lambda_{clip}=0$  & \multicolumn{1}{c|}{28.90}          & 92.23          & \multicolumn{1}{c|}{28.99}              & 92.18              \\ \hline
$\lambda_{identity}=0$    & \multicolumn{1}{c|}{28.74}          & 91.12          & \multicolumn{1}{c|}{28.76}              & 92.01              \\ \hline
\end{tabular}}
\label{tab:ablation}
\end{table}

\section{Discussion and Limitations}
\label{sec:limitations}
We observed only minimal benefits from using CLIP guidance, identity, and CLIP encoding losses in our experiments. Despite experimenting with various metrics to define these losses, such as Cosine, L2, and Contrastive losses, we observed similar behavior. This outcome is likely due to the dominating effect of the adversarial loss. Our discriminator, leveraging the powerful feature extraction capabilities of the pre-trained BiomedCLIP's ViT and MSE loss, can distinguish fake images early in the training process, effectively acting as a regularizer. This regularization may limit the efficacy of CLIP guidance, which can sometimes provide incorrect directions in translation, as shown in earlier studies \cite{sauer2023stylegan,kwon2022diffusion}. Consequently, the impact of CLIP guidance is constrained. Additionally, our model incorporates BiomedCLIP as the multi-modal foundation model to utilize the extensive domain knowledge it gathered during pre-training. Thus, the translation performance of our method is also limited by the contrastive pre-training strategy, where the ViT, focusing on semantically meaningful feature extraction, may suboptimally extract low-level image features. Furthermore, our model may show sensitivity to the captions chosen for the target domain description. Although we tested a range of caption styles, we did not see notable benefits. Therefore, we opted for the simplest combinations employed in BiomedCLIP \cite{zhang2023large}, such as "This is a photo of XX-weighted MRI" and "This is pelvic MRI" or "This is pelvic CT". We leave exploring this aspect further as a future research direction.

Our model employs a generative adversarial network thus, could be fragile to known problems in GAN training like mode collapse. Additionally, GAN training can be unstable and sensitive to hyper-parameter choices, further complicating the training process and potentially hindering the model's effectiveness. 

Our method can be readily adapted to non-medical domains with an expanding range of translations. The primary constraints would be the inherent capacity of the lightweight generator and the domain knowledge embedded in the utilized pre-trained vision-language model. 

\section{Conclusion}
We proposed an unsupervised multi-modal image translation framework employing a generative adversarial network which is empowered with a pre-trained vision-language model. Our framework improves upon the cycle-consistent translation models while enhancing the multi-domain translation performance with a reduced computational budget.

\clearpage 
{\small
\bibliographystyle{ieee_fullname}
\bibliography{main}
}

\end{document}